\documentclass[prl,twocolumn,letterpaper,superscriptaddress]{revtex4}

\usepackage{graphicx}
\usepackage{pslatex}

\begin{document}

\title{Observation of Ultra-high-energy Cosmic Rays\\ 
with the ANITA Balloon-borne Radio Interferometer}

\author{
S.~Hoover$^1$,
J.~Nam$^2$,
P.~W.~Gorham$^3$,
E.~Grashorn$^4$,
P.~Allison$^3$,
S.~W.~Barwick$^5$,
J.~J.~Beatty$^4$,
K.~Belov$^1$, 
D.~Z.~Besson$^6$,
W.~R.~Binns$^7$,
C.~Chen$^8$,
P.~Chen$^{8,11}$,
J.~M.~Clem$^9$,
A.~Connolly$^{10}$,
P.~F.~Dowkontt$^7$,
M.~A.~DuVernois$^{3,12}$, 
R.~C.~Field$^{11}$,
D.~Goldstein$^5$,
A.~G.~Vieregg$^1$,
C.~Hast$^{11}$,
C.~L.~Hebert$^3$,
M.~H.~Israel$^7$,
A.~Javaid$^9$.
J.~Kowalski$^3$,
J.~G.~Learned$^3$,
K.~M.~Liewer$^{13}$,
J.~T.~Link$^{3,14}$,
E.~Lusczek$^{12}$,
S.~Matsuno$^{3}$,
B.~C.~Mercurio$^4$,
C.~Miki$^{3}$,
P.~Mio\v{c}inovi\'c$^{3}$,
C.~J.~Naudet$^{13}$,
J.~Ng$^{11}$,
R.~J.~Nichol$^{10}$,
K.~Palladino$^4$,
K.~Reil$^{11}$,
A.~Romero-Wolf$^3$,
M.~Rosen$^{3}$,
L.~Ruckman$^3$,
D.~Saltzberg$^1$,
D.~Seckel$^9$,
G.~S.~Varner$^{3}$,
D.~Walz$^{11}$,
F.~Wu$^{5}$
}
\vspace{2mm}
\noindent
\affiliation{
Dept. of Physics and Astronomy, Univ. of California, Los Angeles, CA 90095.
$^{2}$Dept. of Physics, Ewha Womans University, Seoul, South Korea.
$^{3}$Dept. of Physics and Astronomy, Univ. of Hawaii, Manoa, HI 96822. 
$^{4}$Dept. of Physics, Ohio State Univ., Columbus, OH 43210. 
$^{5}$Dept. of Physics, Univ. of California, Irvine, CA 92697. 
$^{6}$Dept. of Physics and Astronomy, Univ. of Kansas, Lawrence, KS 66045. 
$^{7}$Dept. of Physics, Washington Univ. in St. Louis, MO 63130. 
$^{8}$Dept. of Physics, Grad. Inst. of Astrophys.,\& Leung Center for 
Cosmology and Particle Astrophysics, National Taiwan University, Taipei, Taiwan.
$^{9}$Dept. of Physics, Univ. of Delaware, Newark, DE 19716. 
$^{10}$Dept. of Physics and Astronomy, University College London, London, United Kingdom. 
$^{11}$SLAC National Accelerator Laboratory, Menlo Park, CA, 94025. 
$^{12}$School of Physics and Astronomy, Univ. of Minnesota, Minneapolis, MN 55455.
$^{13}$Jet Propulsion Laboratory, Pasadena, CA 91109. 
$^{14}$Currently at NASA Goddard Space Flight Center, Greenbelt, MD, 20771.
}

\begin{abstract}
We report the observation of sixteen cosmic ray events of mean energy of $1.5 \times 10^{19}$~eV,
via radio pulses originating from the interaction of the cosmic ray air shower with
the Antarctic geomagnetic field, a process known as geosynchrotron emission. 
We present the first ultra-wideband, far-field measurements
of the radio spectral density of geosynchrotron emission in the range from 300-1000~MHz. 
The emission is 100\% linearly polarized in the plane perpendicular to the 
projected geomagnetic field. Fourteen of our observed events are seen to have a
phase-inversion due to reflection of the radio beam
off the ice surface, and two additional events are seen directly from above the horizon.
\end{abstract}
\maketitle

The origin of ultra-high energy cosmic rays (UHECR) remains a mystery
decades after their discovery~\cite{Nagano00,Auger10}.
Key to the solution will be increased statistics on events of high enough energy
($\geq 3 \times 10^{19}$~eV) to elucidate the endpoint of the UHECR energy spectrum as
seen at Earth.
The primary difficulty is the extreme rarity of events at these energies.
Despite steady progress with experiments such as the 
Pierre Auger Observatory, there remains room for
new methodologies. 
Cosmic rays have been detected for decades via 
impulsive radio geosynchrotron emission~\cite{Jelley65,Porter65,Vernov67,Barker67,Fegan68,
Hazen69,Hazen70,Spencer69,Fegan69,Allan71,Codalema,LOPES,LOPES10}
but until now not in this crucial energy range, which offers the possibility
of pointing the UHECRs back to their sources.  We present data from
the Antarctic Impulsive
Transient Antenna (ANITA)~\cite{ANITA-inst} which represents
the first entry of radio techniques into this energy range.  We find 16
UHECR events, at least 40\% of which are above $10^{19}$~eV, and we show
compelling evidence of their origin as
geosynchrotron emission from cosmic-ray showers. Our results indicate degree-scale precision for 
reconstruction of the UHECR arrival direction, lending strong credence
to efforts to develop radio geosynchrotron detection as a 
competitive method of UHECR particle astronomy.

Geosynchrotron emission arises when the electron-positron particle cascade
initiated by a primary cosmic ray encounters the Lorentz force
in the geomagnetic field. The resulting
acceleration deflects the electrons and positrons and they begin to spiral
in opposite directions around the field lines~\cite{FalckeGorham,Suprun03}. 
In air, the particles' radiation length
is of order 40 g cm$^{-2}$,  a kilometer
or less at the altitudes of air shower maximum development. Particle
trajectories form partial arcs around the field lines before they lose enough 
energy to drop out of the shower. 
The meter-scale longitudinal thickness of the shower particle `pancake' is
comparable to radio wavelengths below several hundred MHz; thus the ensemble behavior of all of the
cascade particles yields forward-beamed synchrotron emission
which is partially or fully coherent in the radio regime.
Therefore, the resulting radio
impulse power grows quadratically with primary particle energy,
and at the highest energies, yields radio pulses that are
detectable at large distances. 
Current systems under development for detection of these radio impulses are 
co-located with cosmic-ray particle detectors on 
the ground to aid in cross-calibration~\cite{Codalema,LOPES,LOPES10}.  
They detect showers with primary energies in the  $10^{17-18}$~eV  range 
because of their limited acceptance.  No such system
has reported a sample of $>10^{19}$~eV UHECR events.

\begin{figure}[htb!]
\includegraphics[width=3.3in]{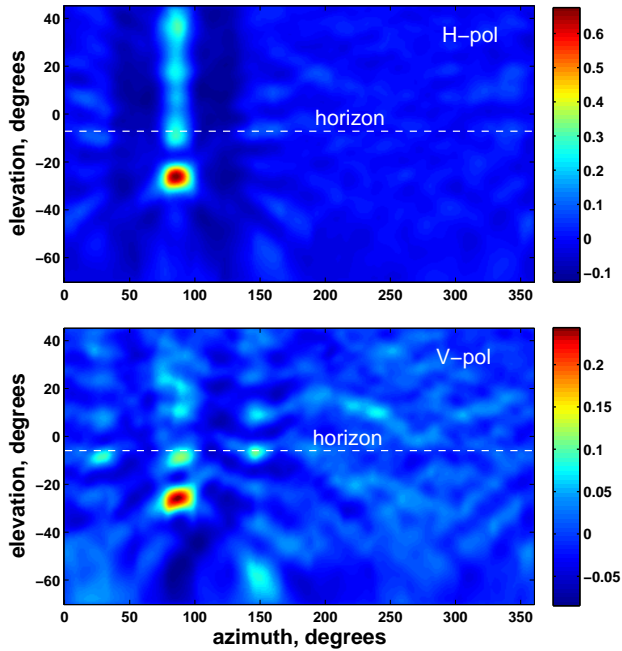}
\caption{An example of interferometric maps of relative correlated intensity 
for both Hpol (top) and Vpol (bottom) from event
3623566 which occurred in a region of Antarctica where the geomagnetic inclination
gave an appreciable Vpol component for the shower radio emission. 
The sidelobes are residuals from the relatively sparse sampling of the 
ANITA interferometer baselines.  Such maps are used to verify the location of the 
emission source on the Antarctic continent, and exclude emission that arises from known
anthropogenic sources.
\label{HVmap}}
\end{figure}

\begin{figure}[htb!]
\includegraphics[width=3.9in]{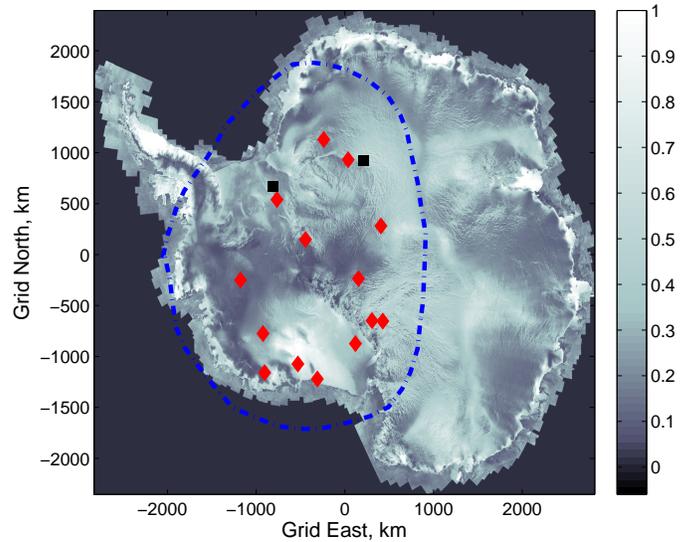}
\caption{Map of locations of detected UHECR events superimposed on
a Radarsat image of relative microwave radar backscatter amplitude of the Antarctic
continent. The red diamonds are the reflected-event locations,
the black squares are the two direct-event locations. The dash-dot line indicates
the limit of ANITA's field-of-view for the flight. Note that the portion
of ANITA's field-of-view that includes the ocean was always covered by
sea-ice during the flight.
\label{icemap}}
\end{figure}

The ANITA long-duration balloon payload is 
launched from Williams Field near McMurdo Station, Antarctica. It
takes advantage of the stratospheric South Polar Vortex to circle the Antarctic continent
at altitudes of 35-37~km while synoptically observing an area of ice of order $1.5$M~km$^2$.
During flight, ANITA records all nanosecond-duration radio impulses over a 
200-1200~MHz radio frequency band.
The threshold is a few times the received power of thermal emission 
from the ice, $\sim10$ picoWatts. The direction of detected signals,
determined by pulse-phase interferometric mapping 
(Fig.~\ref{HVmap},\cite{ANITA-inst}),
is localized to an angular ellipse of $0.3^{\circ} \times 0.8^{\circ}$ 
(elevation $\times$ azimuth) which
is projected back onto the continent to determine the origin of the pulse.
ANITA's mission is the detection of ultra-high energy neutrinos via linearly-polarized 
coherent radio Cherenkov pulses from cascades the neutrinos initiate within the
ice sheets. Virtually all impulsive signals detected during a flight are
of anthropogenic origin, but such events can be rejected with high confidence because
of their association with known human activity, which is carefully monitored in Antarctica. 
For its first flight, during the 
2006-2007 Austral summer, ANITA's trigger system was designed to maximize sensitivity
to linearly polarized radio pulses, but purposely
blinded to the plane of polarization. However, the entire
polarization information -- both vertical and horizontal (Vpol and Hpol) -- was 
recorded for subsequent analysis. Since radio
pulses of neutrino origin strongly favor vertical polarization,
due to the geometric-optics constraints on the radio Cherenkov cone as it refracts through
the ice surface, we used the Hpol information as a sideband test for
our blind neutrino analysis.

\begin{figure}[htb!]
\includegraphics[width=3.15in]{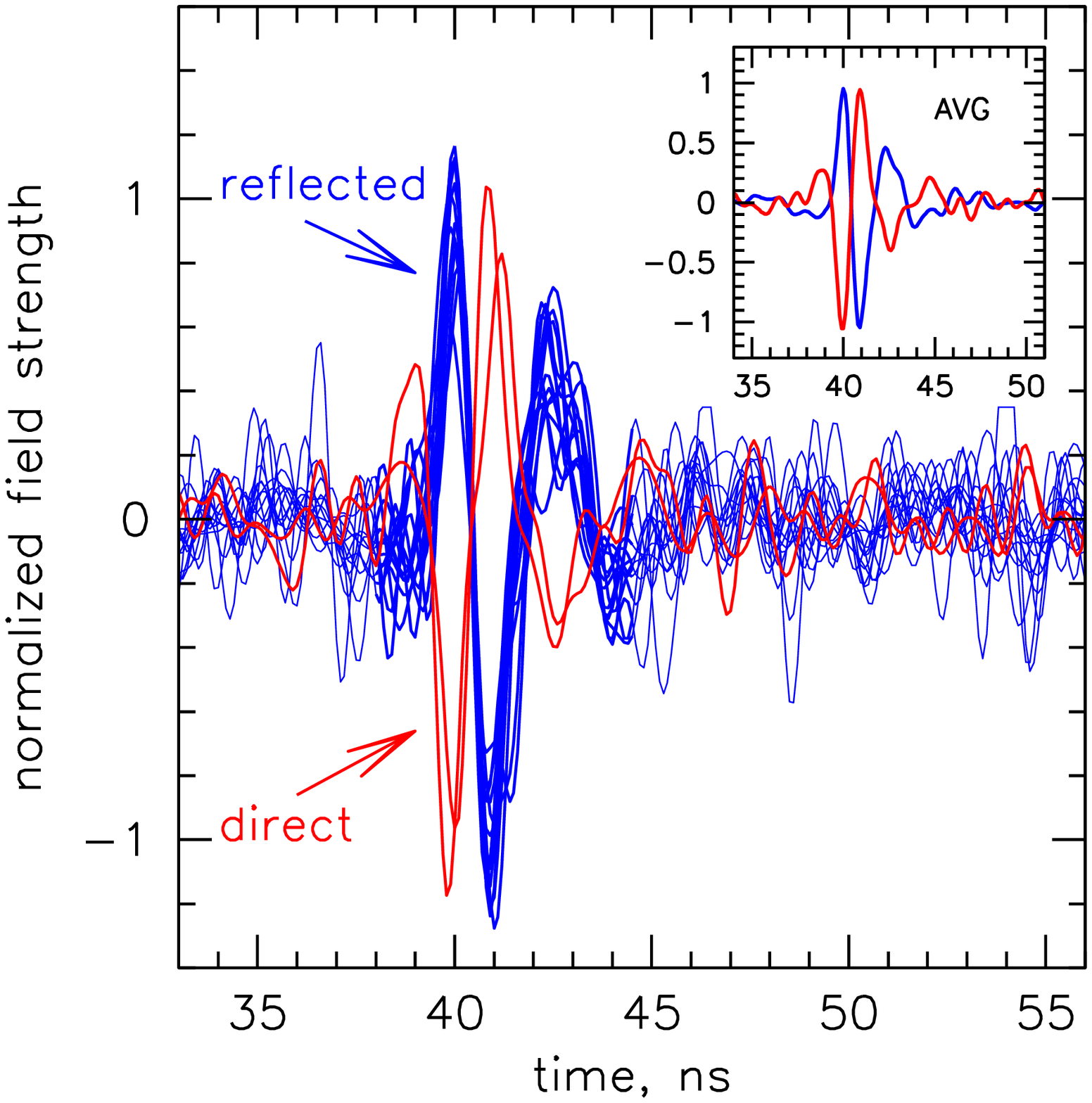}\\
\includegraphics[width=3.15in]{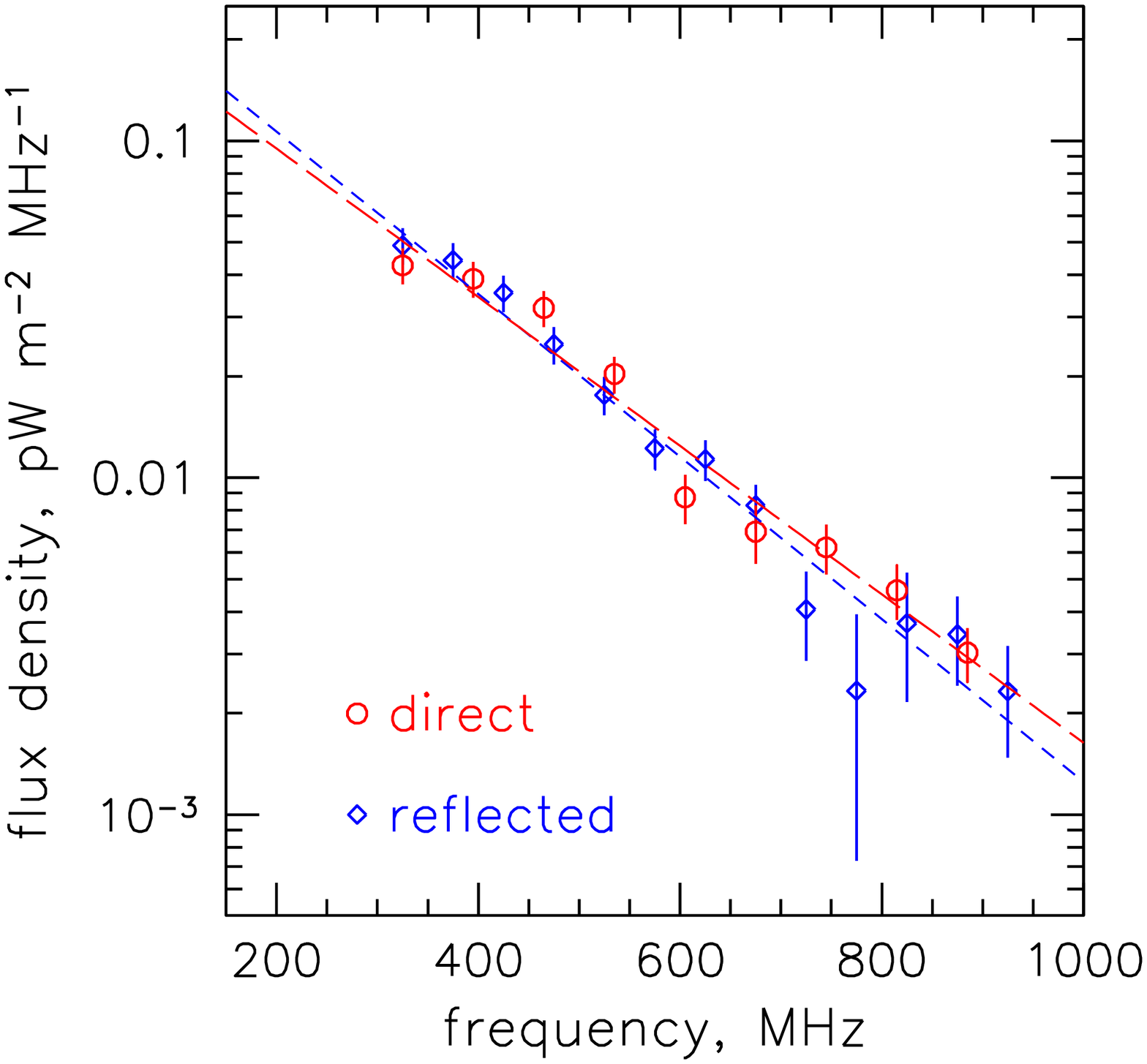}
\caption{Top: overlay of the 16 UHECR event Hpol pulse shapes,
showing the inverted phase for the 14 reflected events (in blue)
compared to the two direct events (in red). 
Inset: Average pulse profile for all events.
Bottom: Flux density for both the averaged direct and averaged reflected events.
In each case the data are consistent with an exponential decrease
with frequency: the fitted coefficients of
decrease with frequency are ($180 \pm 13~{\rm MHz})^{-1}$, and
($197 \pm 15~{\rm MHz})^{-1}$, consistent with each other within fit errors.
Errors at low frequency (high SNR) are primarily due to systematic
uncertainty in the antenna gains, and to thermal noise statistics
at higher frequencies.
\label{pulse_spectra}}
\end{figure}

\begin{figure}[htb!]
\includegraphics[width=3.3in]{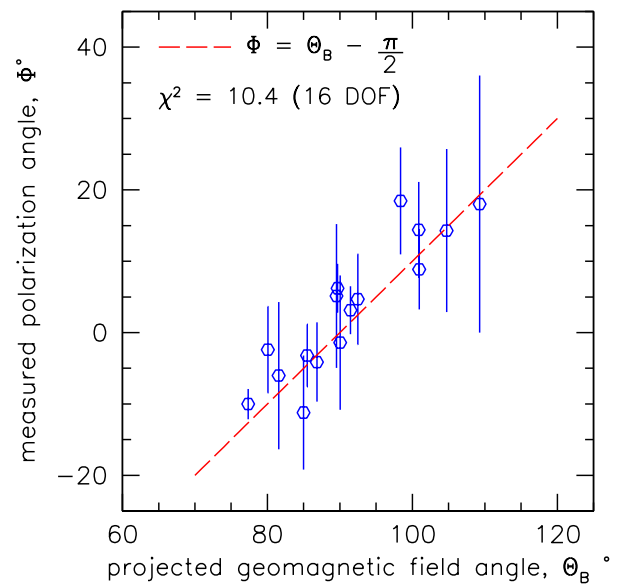}
\caption{Plane of polarization of UHECR events compared to the
angle of the magnetic field local to the event, with the
red line indicating the expectation for the Lorentz force.
The reflected events are corrected for their surface Fresnel 
coefficients, and angles are measured from the horizontal.\label{polangle}}
\end{figure}

Our results were surprising: 
while the neutrino analysis (Vpol) gave a null result, a statistically
significant sample of 6 Hpol events was found initially~\cite{ANITA-1},
and a more sensitive analysis now yields 16.
These events are randomly distributed around ANITA's flight path (Fig.~\ref{icemap}), 
uncorrelated in location
to human activity or to each other, but closely correlated to each other in their radio 
pulse profile and frequency spectrum (Fig.~\ref{pulse_spectra}, Top). 
Their measured planes of polarization are found in every
case to be perpendicular to the 
local geomagnetic field (Fig.~\ref{polangle}), as
expected from geosynchrotron radiation. With two
exceptions, the events reconstruct to locations on the surface of the ice; the
two exceptional cases have directional origins above the horizon, but below the
horizontal (from stratospheric balloon altitudes the horizon is about $6^{\circ}$
below the horizontal). Earth-orbiting satellites are excluded 
as a possible source since the nanosecond radio temporal coherence
observed is impossible to retain for signals that propagate through the ionospheric
plasma, which is highly dispersive in our frequency regime.
The fourteen below-horizon events are inverted compared to the two above-horizon events,
as expected for specular reflection (Fig.~\ref{pulse_spectra}, Top). 
From these observations we conclude that ANITA detects a signal, seen in 
most cases in reflection from the ice sheet surface,
which originates in the earth's atmosphere and which involves electrical current
accelerating transverse to the geomagnetic field. Such observations are in
every way consistent with predictions of geosynchrotron emission from
cosmic-ray air showers. In addition, the inherent spectral and time-domain 
similarity of our radio pulses,
as well as their robust correlation to geomagnetic parameters, suggests that ANITA's observations,
which are at much greater distance and higher frequency than prior and current
air-shower geosynchrotron observations, are less susceptible to near-field 
fluctuations of radio strength and plane of polarization. Such issues have
been problematic in this field throughout most of its history.

Our data represent the first broadband measurements of geosynchrotron
emission in the UHF frequency range.
The average observed radio-frequency spectral flux density of the 
above- and below-horizon events, shown in Fig.~\ref{pulse_spectra} (Bottom)
is consistent with an exponential decrease with frequency.
 The lack of any statistically significant difference in 
the spectra for the direct and reflected events indicates 
that ice roughness is unimportant for the average surface reflection.
To estimate the electric field amplitude at the source of these emissions, we model the
surface reflection using standard physical-optics treatments
developed for synthetic-aperture radar analysis.
Such models use self-affine fractal surface parameters~\cite{Shepard99} and Huygens-Fresnel
integration over the specular reflection region to estimate both amplitude
loss and phase distortion from residual slopes or roughness. In our case, we used 
digital-elevation models from Radarsat~\cite{Radarsat} to estimate surface parameters
for each of the event reflection points, known to a few km precision. 
In most cases the surface parameters
are found to be smooth, yielding only modest effects on the reflection
amplitude; in a minority of the events, surface parameters were estimated to be
rougher, but still within the quarter-wave-rms Rayleigh criterion for coherent
reflection~\cite{Radiobook}.  Fresnel reflection coefficients were determined using
a mean near-surface index of refraction of $n = 1.33$, typical of Antarctic firn.

\begin{figure}[htb!]
\includegraphics[width=3.3in]{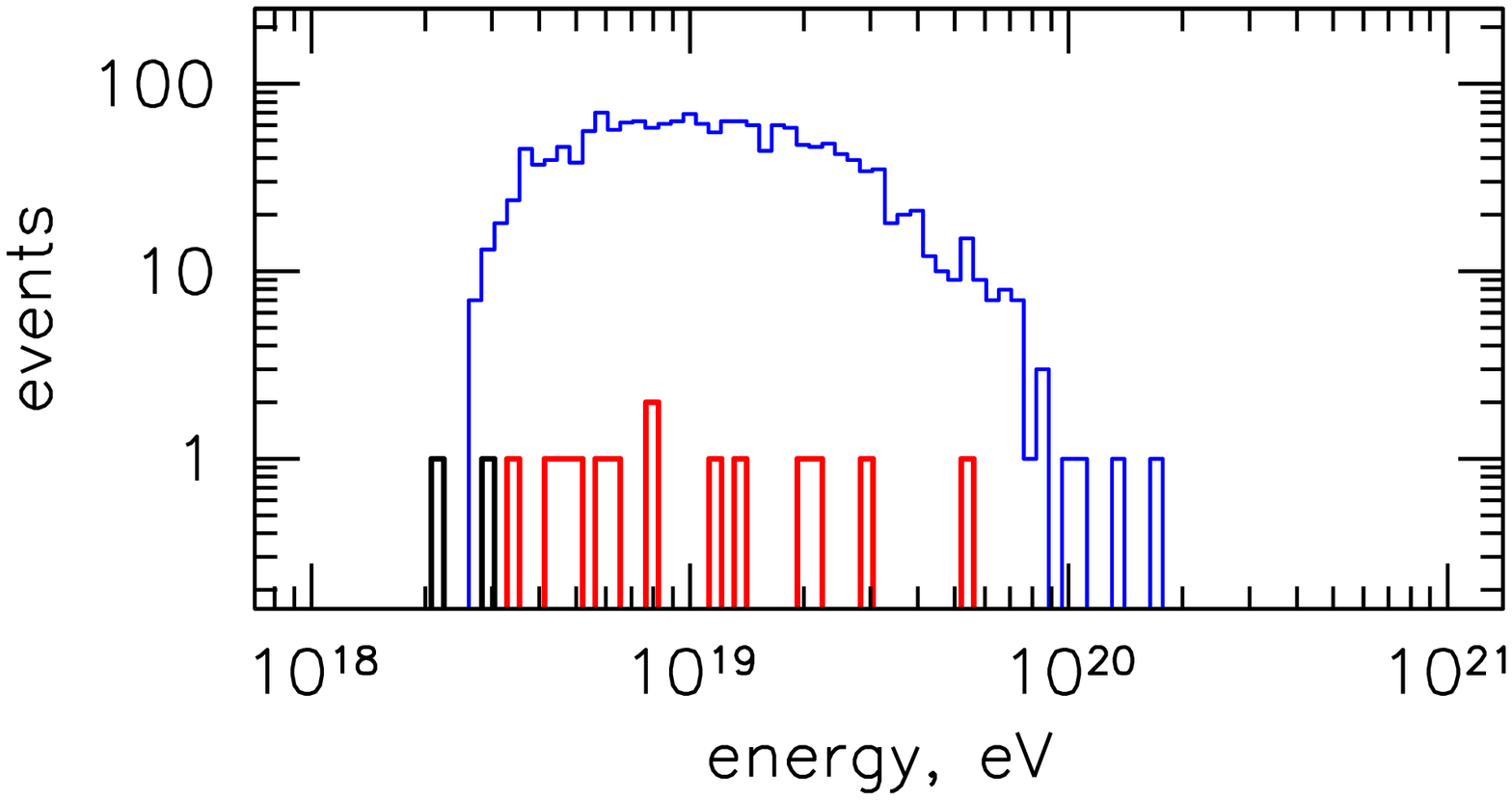}\\
\includegraphics[width=3.3in]{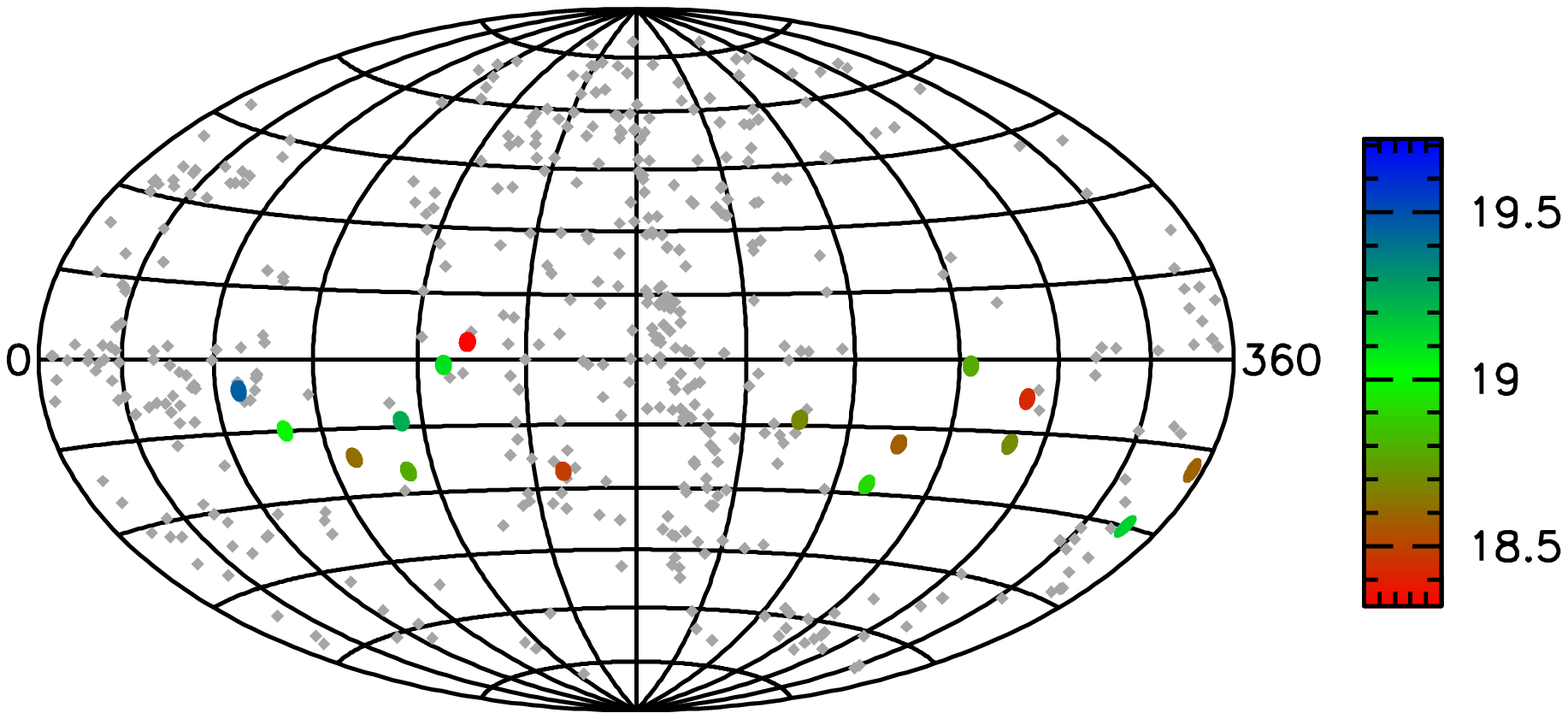}
\caption{Top: energies of detected UHECR events, with reflected
events in red, direct (above-horizon) events in black, and the
simulated event sample (reflected events only) shown in blue. 
Bottom: Map in celestial ($\alpha,\delta$) coordinates of the ANITA events (circles) with
2.0 degree radii, and nearby AGN (grey diamonds) from the V\'{e}ron-Cetty catalog. 
The approximate energy for each event is color coded by the log$_{10}$ of the
estimated event energy. ANITA's exposure is approximately uniform across
the band $5^{\circ}>\delta >-30^{\circ}$.\label{energy+map}}
\end{figure}

To estimate the primary energy for the observed events, we 
used two independent approaches that determine the amplitude 
of the radio emission and the mean angular offset of the observed 
events.  One approach is based on current air-shower geosynchrotron 
radio emission simulations developed for surface arrays~\cite{LOPES,HF05,HEU07}, 
and the second approach is based on data-driven maximum-likelihood 
modeling in which a small set of parameters of a semi-empirical model 
were iteratively fit to the observed characteristics  
and total number of the events, given the known UHECR energy spectrum~\cite{Auger10}.
The former method had the advantage of extensive work done to develop full-scale air shower
Monte Carlo simulations for such radio emission; however, the simulations are not
directly relevant to the very different geometry and higher frequency range
of ANITA's observations, which
are in the far-field compared to most ground array observations, and which also
involve showers at much larger zenith angles than ground arrays usually observe.
The latter data-driven approach used physically-motivated parameterizations to capture the
radio emission characteristics. The resulting constraints imposed 
by the data, including
amplitude, phase, and frequency-spectral content for all 16 events, were found to
be effective in fitting both the primary energy
and observed angular distribution of the events. Using the ground-based geosynchrotron models we found
no self-consistent solution for the event energy and mean angular offset,
and we conclude that ANITA's observations are in tension with the current
ground-based simulations. Our data-driven approach converged on a
solution which gave estimated event energies as shown in Fig.~\ref{energy+map} along with
an overlay of the histogram of the energy distribution of simulated events 
seen in reflection. The implications of the data-driven solution are that the
RF signals from these highly inclined, distant showers are significantly stronger than 
predicted by current geosynchrotron models.
The mean energy of the ensemble of reflected events is estimated to be 
$1.5 \pm 0.4 (stat) ^{+2.0}_{-0.3} (sys) \times 10^{19}$~eV,  
approaching the threshold of the Greisen-Zatsepin-Kuzmin (GZK)
cutoff~\cite{G,ZK}, which marks the beginning of the absorption edge of
UHECRs against the cosmic microwave background radiation. The large asymmetry in
the systematic uncertainty is due to the uncertainty in the angular offset,
which tends to strongly bias toward underestimating the event energy in our
models. For the direct events,
the mean energy is lower due to stronger direct signals, but the acceptance -- 
limited to a narrow angular band around the horizon -- is also much lower.

Based on our data-driven semi-empirical approach, 
we estimate a mean angle of observation relative to the true shower axis 
of $(1.5 \pm 0.5) ^{\circ}$. 
This angular precision is comparable to that of ground-based cosmic-ray observatories,
and adequate to allow us to map these events back to the sky.
The final error circle is $2^{\circ}$ in diameter after convolving with
angular reconstruction precision and
the modest tilts of each event locale, determined from Radarsat images at
200~m resolution~\cite{Radarsat}.
The resulting map is shown in Fig.~\ref{energy+map}.
Our event positions are uncorrelated to the sky positions of the Auger Observatory
UHECR events, and the ensemble is also uncorrelated to AGN in the nearby universe. This is expected 
for events in this energy range where intergalactic magnetic deflection is significant. 
While our sample of UHE events is significantly
smaller than the current totals for the Auger Observatory~\cite{Auger10}, according to our models
the acceptance of this method of UHE detection continues to increase at high energies, 
even beyond $10^{20}$~eV, whereas the acceptance of
all ground-based UHECR observatories saturate well before this.
Estimates from our simulations indicate that, 
after optimization for UHECR observation, 
a new 30 day flight of ANITA could detect
a total of several hundred geosynchrotron events, with 60-80 above $10^{19}$~eV, and
$\sim 10$  above the nominal GZK cutoff energy. 
We conclude that a balloon-borne observatory is viable at the highest 
cosmic-ray energies, and if the fidelity of models of the geosynchrotron process continues to improve
at the rate it has in recent years, such an approach will be able to further elucidate
possible correlations in cosmic-ray origin directions as well as 
the shape of the endpoint of the UHECR energy spectrum. 

We are grateful to NASA, the US National Science Foundation, the US Dept. of
Energy and the Columbia Scientific
Balloon Facility for their generous support of these efforts.


\end{document}